\documentclass[epj]{svjour}
\usepackage{graphics}
\usepackage{graphicx}
\begin{document}
\title{Theoretical Summary}
\subtitle{The First International Conference on Hard and Electromagnetic Probes \\ 
in Relativistic Nuclear Collisions}
\author{Dmitri Kharzeev}
%
%
\institute{Nuclear Theory Group\\
Physics Department \\ Brookhaven National Laboratory \\ 
Upton, NY 11973, USA}
\date{Received: date / Revised version: date}
%
\abstract{
This is an attempt to summarize the theoretical talks given at the 
First International Conference "Hard Probes '04", dedicated 
to the study of the properties of quark-gluon matter and its diagnostics 
with the hard processes.
 The talk covers the 
following topics: the structure of quark-gluon matter at finite temperature; the theory of nuclear wave functions at small Bjorken $x$; 
the propagation of jets, heavy quarkonia and heavy quarks 
through the dense QCD matter.
} 
\maketitle
\section{Introduction}
\label{intro}
The venue of this Conference -- the small town of Ericeira on the Atlantic coast near Lisbon -- 
is both spectacular and symbolic. We are at the western end of Europe, 
a place which calls to mind the history of how the New World was discovered. 
At the end of 15th century, nothing was known yet about the new lands hidden 
by the extensive ocean. Yet, the discoveries were already anticipated by some, and in 
1494 the Pope divided the world to be discovered between Portugal and Spain, in the Treaty 
of Tordesillas. The sharp, straight boundary extended from North to South and divided what was at the time believed to be an empty 
ocean; less than 10 years later, South America had been discovered. The subsequent exploration of the New World made the shape of the boundary 
much more complex, and the subsequent developments eventually made it irrelevant altogether. What lessons can be learnt from this story? 
In my opinion, there are at least three:
\vskip0.2cm

i) the less we know, the sharper are the boundaries;

ii) sharp boundaries do not last long --

iii) they disappear with the advance of knowledge. 

\vskip0.3cm

As Helmut Satz reminded us in his opening talk, this conference grew out of 
the "Hard Probe caf\'{e}", which had its first meeting at CERN, in 1994 -- five centuries 
after the Treaty of Tordesillas. The discoveries at the high energy density and small x 
frontiers were widely anticipated, and the boundaries on the QCD maps were still very sharp. 
Regarding the statistical properties of QCD, most of us expected to see the weakly interacting quark-gluon gas just above the deconfinement 
temperature, although the lattice data already at that time indicated large deviations 
from the ideal gas behavior  -- see \cite{fkarsch}. As for the behavior of QCD 
at high energies (or small Bjorken $x$), it was widely believed that the transition from 
"soft" to "hard" regimes happens at some typical scale $Q_0 \sim 1\div 2$ GeV, which 
does not depend on the energy, even though the idea of parton saturation \cite{GLR,Mueller:1985wy,Blaizot:1987nc} was already 
known and the related classical gluon field approach \cite{MV} had just been developed.  

The experimental heavy ion program at CERN SPS was blooming, and the great potential 
held by the hard probes had already been made clear by the discovery of 
$J/\psi$ suppression \cite{NA50} predicted by Matsui and Satz \cite{MS} (even though the
interpretation of the data was a subject of intense discussions). 
The low--mass dilepton enhancement \cite{Agakishiev:1995xb} was observed shortly afterwards and attracted a lot of attention 
as a potential signature of chiral symmetry restoration, and  
Drell--Yan pair production 
proved to be very useful as the baseline. However, high transverse momentum hadrons, let alone jets, 
were very rare at the SPS energy ($\sqrt{s} \leq 20$ GeV per nucleon pair). 

The new millenium brought RHIC  -- and with it, the era of hard probes in relativistic 
heavy ion physics has begun.   At this Conference, we have heard about the amazing progress 
made in the experimental study of hard processes in recent years; the excellent overviews of the current situation  were made at this conference \cite{Drees,Harris:2005vt,Jacobs:2005pk,Scomparin,Specht}. 

So what have we learnt so far from this wealth of experimental information, and what do we still need to know? 
In what follows below, I try to address these questions from the 
theorists' point of view, based on the talks given at the Conference and 
on some of my own prejudices. 
The space limits prevent me from describing all of the reported exciting developments, so instead of presenting a catalogue of the given 
talks I will concentrate on a few selected topics.

\section{Quark-gluon matter at high temperature}
\label{sec:1}

\subsection{Strongly coupled quark-gluon plasma: a surprise?}
\label{subsec:12}

For years, we have been expecting that at "sufficiently high" temperature $T$ the QCD matter 
will become an "almost" ideal gas of quarks and gluons. Indeed, a typical inter-particle distance 
in this matter is $\sim 1/T$, and the asymptotic freedom tells us that the interactions at short 
distances are weak. We still hold this expectation, but the data from RHIC tell us that "sufficiently high" temperatures appear 
beyond the reach of the current, and perhaps future, experiments: at all  
accessible temperatures  the QCD matter behaves quite differently from an ideal gas, as emphasized at this Conference by 
E. Shuryak \cite{shuryak} and others. The dynamics 
of the quark--gluon plasma is thus much more rich and interesting, and we have to develop new 
methods to understand it.  

\begin{figure} 
\resizebox{0.50\textwidth}{!}{%
\includegraphics[]{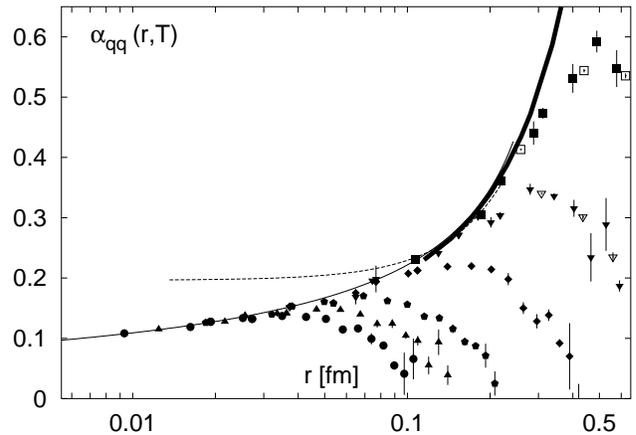}
}
\caption{QCD running coupling for temperatures above the deconfinement transition; the sets of points 
correspond to (going down) $ T/T_c = 1.05; 1.1; 1.2; 1.3; 1.5; 1.6; 3.0; 6.0; 9.0; 12.0 $; the solid line is for $T=0$.
 From \cite{Kaczmarek:2004gv,fkarsch}
}
\label{eff_as}
\end{figure}

In fact, as discussed at the Conference by F. Karsch \cite{fkarsch}, there have been numerous indications 
from lattice QCD that even above the deconfinement transition the interactions among quarks and gluons 
remain strong.  A particularly telling piece of evidence from the lattice calculations is presented in Fig.\ref{eff_as}, 
which shows the behavior of the QCD running constant as a function of distance for different 
temperatures. At $T=0$, one observes the celebrated property of asymptotic freedom, or anti--screening of the 
color charge. Above the deconfinement temperature, the strong force gets screened -- in agreement 
with the qualitative picture in which the range of the interaction is reduced because the exchanged gluons can 
scatter off the heat bath of deconfined thermal quarks and gluons. 
However, at experimentally accessible temperatures the screening develops at relatively large distances, 
at which the coupling constant is quite large. We are thus definitely dealing with a deconfined quark-gluon plasma, 
in which the long--range confining interactions are screened, but the residual non--perturbative effects are still 
strong. 

This property of the observed quark-gluon plasma makes the traditional quasi--particle description of its excitations 
questionable, as discussed by J.-P. Blaizot \cite{Blaizot:2005mj} and K. Rajagopal \cite{rajagopal}, and one has to re--identify 
the appropriate degrees of freedom. Blaizot pointed out in particular the experimental implications of this problem for the dilepton production rates. 
Rajagopal also discussed the corresponding problem in the theory of cold quark matter, described as a color super-conductor, and described 
the applications to the physics of neutron stars. The ways to test the structure of the quark-gluon plasma 
in lattice simulations and in experiment include the study of fluctuations, as discussed by R. Gavai \cite{gavai} 
and various transport coefficients, including viscosity \cite{shuryak,Hatsuda}.

\subsection{Quarkonium suppression in a strongly coupled Quark-Gluon Plasma}

As pointed out long time ago by Matsui and Satz \cite{MS}, the study of heavy quarkonia in hot QCD matter allows 
to test the persistence of confining interactions. Indeed, this is probably the closest one can get in experiment to measuring the 
order parameter of the deconfinement -- the large distance limit of the correlation function of the Polyakov loops, 
which measures the interaction energy of the separated heavy quark and antiquark \cite{McLerran:1980pk}. 
Therefore, if some residual non-perturbative interactions are present above $T_c$, they may manifest themselves 
in the spectra of heavy quarkonia.   

Very interesting lattice results on this issue have been presented at the Conference 
by T. Hatsuda \cite{Hatsuda}, P. Petreczky \cite{Petreczky:2005bd}, K. Petrov \cite{Petrov:2005sd}, S. Digal \cite{Digal:2005ht}, O.~Kaczmarek and F.~Zantow 
\cite{Kaczmarek:2005uv}. All of them point towards the survival of some of the bound charmonium states in the deconfined phase, 
which is consistent with the large screening radius of Fig. \ref{eff_as}. 
\begin{figure} 
\resizebox{0.50\textwidth}{!}{%
\includegraphics[]{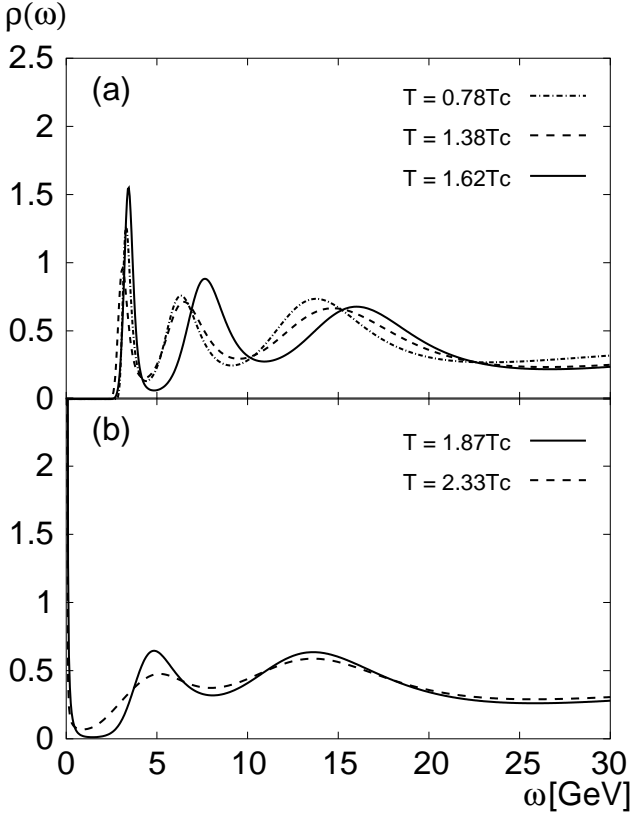}
}
\caption{Thermal vector $\bar{c}c$ spectral functions extracted from the maximal entropy method analysis of the quenched lattice QCD calculations; from \cite{Hatsuda,Asakawa:2003re}, see also \cite{Datta:2003ww,fkarsch}
}
\label{corrf}
\end{figure}
There are two basic ways of accessing the information 
about charmonia on the lattice: one is to measure the correlation function of the $\bar{c}c$ current and to reconstruct 
the corresponding spectral function, another is to compute the effective potential between static sources and to use it in 
the Schroedinger equation for the bound states. 

Each of these methods has advantages and difficulties, so they are complementary 
to each other: in the spectral 
function method, one does not have to rely on a potential model, but a reconstruction of the quarkonium spectrum from the data 
has a limited precision. The effective potential approach provides a precise information on the spectrum, but the validity 
of the potential model in a heat bath and a treatment of the coupling between the color-singlet and octet components raise some questions. 

\begin{figure} 
\resizebox{0.50\textwidth}{!}{%
\includegraphics[]{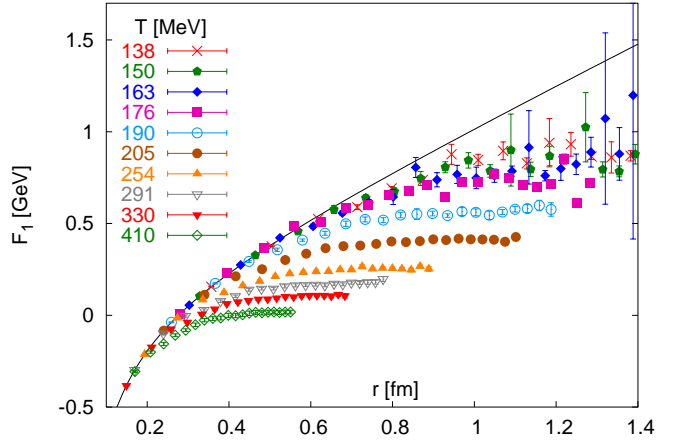}
}
\caption{The color singlet free energy in three flavor lattice QCD; the solid line is the $T=0$ singlet potential. 
>From  \cite{Petreczky:2004pz,Petreczky:2005bd}.}
\label{pot}
\end{figure}

A representative result for the shape of the quarkonium spectral function as extracted from the lattice vector $\bar{c}c$ correlation functions 
(the $J/\psi$ channel) with the help of a MEM (Maximal Entropy Method) approach is shown in Fig. \ref{corrf}.  
One can clearly see that up to temperatures of about $T \sim 2\ T_c$ the peak corresponding to the bound $J/\psi$ state 
still survives in the spectrum. Moreover, in this temperature range little, if any, change in the mass of $J/\psi$ is observed. 
The effective potential method basing on the lattice results shown in Fig. \ref{pot} leads to 
similar conclusions -- the remnants of the confining interaction ("short strings" ?) still exist in the vicinity of the deconfinement phase 
transition and can support bound states. An interesting analysis aimed at linking the spectral function and potential approaches 
was presented at the Conference by A. Mocsy \cite{Mocsy:2004bv}.

Do these lattice results imply that no $J/\psi$ suppression from quark-gluon plasma should be seen in experiment? In my opinion, 
the answer to this question is "no": even if a quarkonium exists as a bound state, it can still be dissociated by the impact of hard 
deconfined gluons  \cite{Kharzeev:1994pz}, in a process analogous to photo--effect \cite{Shuryak:1978ij}. The relative importance 
of the Debye screening and "gluo--effect" processes is governed by the ratio of quarkonium binding energy $\Delta E$ to the temperature of 
the plasma $T$ \cite{Kharzeev:1995ju,Kharzeev:1996se}: 
\begin{equation}
\Gamma(T) = {\Delta E(T) \over T},  
\end{equation}
where the binding energy depends on the temperature due to Debye screening. In the weakly coupled plasma $\Gamma \ll 1$, 
and the heavy quark bound state simply falls apart with the rate 
\begin{equation}
R = {4 \over L} \ \sqrt{{T \over \pi M_Q}},
\end{equation} 
($L$ is the size of quarkonium, $M_Q$ -- the heavy quark mass) which is the classical high temperature, weak coupling limit of the thermal activation rate. On the other hand, in the strongly coupled 
case of $\Gamma \gg 1$, quarkonium is tightly bound, and the binding energy threshold has to be overcome by the absorption of hard 
deconfined gluons from the heat bath. In this regime, the heavy quark bound states are quasi--stable, but the dissociation rate is quite large and can lead to a significant quarkonium suppression \cite{Xu:1995eb}.

At the Conference, the fate of heavy quarkonium in the medium was further discussed by D. Blaschke \cite{Blaschke:2005jg}, 
R. Rapp  \cite{Rapp:2005rr}, and R. Thews \cite{Thews:2005fs}. The latter talks discussed in particular the possibility 
to create additional quarkonia by recombination of heavy quarks and anti-quarks.  In particular, it was shown  \cite{Thews:2005fs} that 
recombination of heavy quarks leads to a sizable narrowing of the rapidity distribution of $J/\psi$'s in $Au-Au$ collisions at RHIC; 
a high statistics experimental measurement of this distribution can thus help to extract the contribution of this mechanism, or to put an upper bound on it. 

Quarkonium suppression in the percolation approach to deconfinement was discussed by M. Nardi \cite{nardi}; the signature of the 
percolation phase transition in this case is a peculiar centrality and mass number dependencies of the $J/\psi$ survival probabilities, 
which are consistent with the existing NA50, NA60 and PHENIX data. The transverse momentum dependence of the $J/\psi$ suppression 
in this picture still remains an interesting open problem  \cite{Kharzeev:1997ry}. 

\begin{figure} 
\resizebox{0.50\textwidth}{!}{%
\includegraphics[]{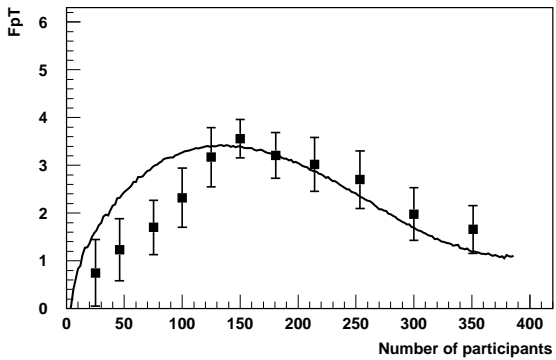}
}
\caption{Fluctuations in the transverse momentum in $Au-Au$ collisions at RHIC at $\sqrt{s} = 200$ GeV; from \cite{Pajares:2005kk}. The data from PHENIX \cite{phenix-fluct} 
is compared to the calculation based on the percolation picture.}
\label{fluct}
\end{figure}

Percolation of strings as a description of deconfinement 
was extensively discussed by J. Dias de Deus \cite{dias} and C. Pajares \cite{Pajares:2005kk}. It was pointed out that the percolation 
approach in particular naturally leads to the observed fluctuations in the transverse momentum (see Fig. \ref{fluct}) and the universal form of the transverse mass distribution of hadrons in nuclear collisions, similar to 
the one arising from the color glass condensate \cite{Schaffner-Bielich:2001qj}. This brings us to the next topic which became 
one of the focal points of the Conference -- the theory of nuclear wave functions on the light cone, at small Bjorken $x$.

\section{High density gluon matter at small $x$}

\subsection{"Just a change of the reference frame?"}

Recent years have seen an impressive progress in the understanding of nuclear wave functions at small 
Bjorken $x$. What makes this problem interesting? After all, nothing changes if we look at the nucleus 
in a different reference frame, where it is boosted to high momentum -- or so it seems at first glance. 
But we have to remember that in quantum theory the operator of the number of particles does not commute 
with the operator of Lorentz boost, and so in general a mere change of the reference frame will change 
the measured number of particles in the system.  

This is certainly the situation in QCD, where the boost 
is accompanied by the evolution of a hadron or nuclear structure function, which leads to a rapid $\sim 1/x^{\lambda}$ growth of the number 
of gluons and quarks at small $x$. Because the boost also leads to the 
Lorentz compression of the nucleus, 
and because the Froissart bound does not allow the area of the nucleus to grow faster than $\sim \ln^2(1/x)$, 
at sufficiently small $x$ and/or large mass number of the nucleus $A$ the density of partons in the transverse 
plane becomes large and they can recombine \cite{GLR,Mueller:1985wy,Blaizot:1987nc}; when the 
occupation number becomes $\sim 1/\alpha_s$, the system can be described as a semi-classical gluon field 
\cite{MV}. A broad overview of the semi--classical Color Glass Condensate approach to nuclear wave functions and to the heavy ion collisions has been presented by R. Venugopalan \cite{raju}. 

\subsection{In search of the ultimate evolution equation}

Once the density of partons becomes large, the non--linear effects in the parton evolution become important. 
The quantum processes of parton splitting and recombination in this regime occur in the background of the strong classical field. The general evolution equation in this case still has to be found, and the progress in this direction 
has been discussed at the Conference by J. Bartels \cite{bartels}, E. Iancu \cite{iancu} and A.H. Mueller. 

A general introduction into the problem of non-linear evolution equations and the underlying physics 
was given by Mueller, who also discussed the limits of validity of the existing approaches. Iancu in particular 
discussed the role of rare fluctuations in hadron wave functions which are not captured by the mean--field 
equation of Balitsky \cite{Balitsky:1995ub} and Kovchegov \cite{Kovchegov:1999yj}. 

One of the important problems of the perturbative QCD approach to high energy scattering 
emphasized by Bartels is the following: in the impact parameter $b$ space, perturbation theory always 
predicts the amplitudes which fall off as inverse powers $ (1 / b)^n$ at large $b$. This is because 
there is no mass gap for the gluon excitations in perturbation theory. On the other hand, in the physical 
world there are no massless hadronic excitations -- pions, as the Goldstone bosons of the spontaneously broken 
chiral symmetry, are the lightest ones, but their masses $m^2_{\pi} \sim m_q$ do not vanish because of the finite light quark 
masses $m_q \neq 0$. Therefore, high energy hadronic scattering amplitudes must fall off exponentially at large 
impact parameters, not slower than $\sim \exp(-2 m_{\pi} b)$ -- coupled with 
the fact that at 
fixed impact parameter the growth of the amplitude is bounded by a power of energy $s$, this leads to the Froissart bound on the total cross sections. Because of the diffusion to large distances in high energy evolution, one 
is forced to consider the influence of the mass gap on the scattering amplitudes. 

\subsection{Probing the Color Glass Condensate} 

Since the growth of parton distributions in the wave function of a nucleus $A$ at small $x$ is tempered by the non-linear 
effects, the rescaled by $A$ number of partons in a heavy nucleus is smaller than in a proton. 
This parton deficit in a heavy nucleus is a quantum effect, which has to manifest itself 
at sufficiently small $x$, when the longitudinal phase space $\sim \ln(1/x)$ for the emitted gluons is large enough  
to compensate the smallness of the coupling, $\alpha_s \ln(1/x) \sim 1$. Indeed, at the classical level 
the total number of partons in a nucleus $A$ is equal to the rescaled number of partons in a nucleons, 
but they are re-distributed in the transverse momentum which leads to the Cronin effect in nuclear 
cross sections. 

The number of partons in the nuclear wave functions can be measured in hard $p(d)A$ scattering 
processes at small $x$; at RHIC this corresponds to the forward rapidity region (the deuteron fragmentation region). Therefore one arrives to the prediction that the cross sections of hard $dA$ scattering in the 
forward rapidity region should be suppressed relative to the $NN$ ones. The physics of this phenomenon 
has been extensively discussed at the Conference by R. Baier \cite{baier},
B. Gay Ducati \cite{gay}, J. Jalilian-Marian \cite{jalilian}, 
J. Milhano and C. Salgado \cite{Albacete:2005ef}, D. Triantafyllopoulos \cite{Triantafyllopoulos:2005eh} and K. Tuchin \cite{Tuchin:2005ky}.  

\begin{figure} 
\resizebox{0.50\textwidth}{!}{%
\includegraphics[]{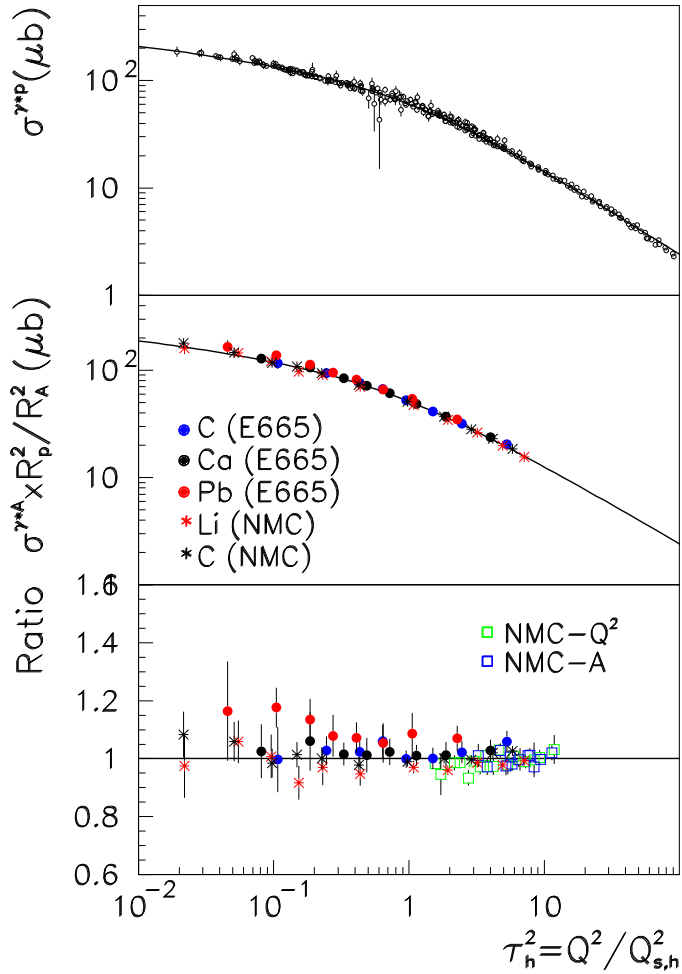}
}
\caption{Geometric scaling for $\gamma^* p$ scattering (upper panel), $\gamma^* A$ (middle panel), 
and the ratio of the $\gamma^* A$ data over the theoretical predictions based on the saturation picture 
(lower panel); from  \cite{Albacete:2005ef}.}

\label{geom}
\end{figure}

\begin{figure} 
\resizebox{0.50\textwidth}{!}{%

\includegraphics[]{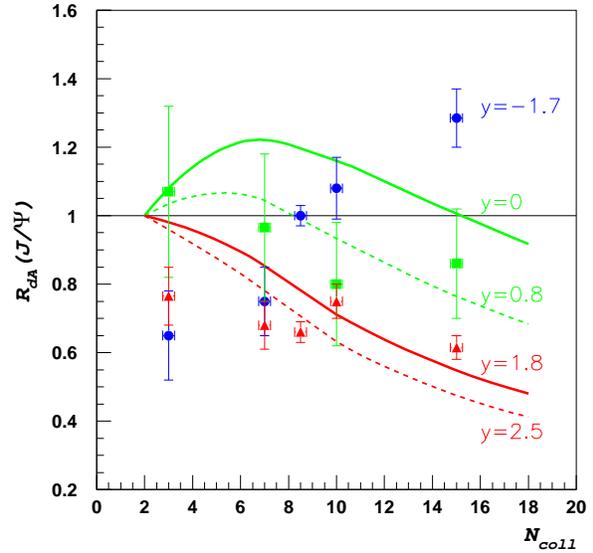}
}
\caption{The nuclear modification factor in the production of $J/\psi$ in $dAu$ collisions at RHIC energy of 
$\sqrt{s} = 200$ GeV. Theoretical calculations based on the color glass condensate picture are compared 
to the preliminary data from PHENIX \cite{phenix-psi}; from  \cite{Tuchin:2005ky}.}

\label{psi-k}
\end{figure}

Jalilian-Marian \cite{jalilian} presented a clear introduction to the problem, and discussed the effects 
of quantum evolution in the color glass condensate on the production of hadrons, dileptons and photons 
at forward rapidities. Dilepton and photon production at forward rapidities have also been the topic of talks 
given by R. Baier \cite{baier} and Gay Ducati \cite{gay}. Baier in particular has 
demonstrated the potential of these probes 
for understanding the nuclear gluon distributions at small $x$. 
Salgado  \cite{Albacete:2005ef} has shown that the saturation picture leads to a consistent description of the small $x$ 
data on deep-inelastic scattering off both protons and nuclei, see Fig. \ref{geom}. He argued that 
this picture also allows to describe the data on hadron multiplicites at RHIC. Triantafyllopoulos \cite{Triantafyllopoulos:2005eh} discussed the transition from the classical to quantum regimes in $pA$ scattering, and the 
evolution and disappearance of the Cronin peak with rapidity. Tuchin \cite{Tuchin:2005ky} presented results on the influence 
of the color glass condensate on the production of charmed quarks and charmonia. In the latter case, he found 
an interesting effect of nuclear $J/\psi$ enhancement in a certain window in rapidity, see Fig.   \ref{psi-k}.

 Much of the existing theoretical analysis is based on the method of $k_T$ factorization. The limitations 
 of this approach were examined by H. Fujii and 
 F. Gelis \cite{Fujii:2005rm} using an example of heavy quark and quarkonium production.
 
 Theoretical approaches currently used for the description of $pA$ collisions were discussed by J. Qiu \cite{qiu}; 
 he analyzed the contributions of higher twist effects resulting from coherent multiple scattering, and their 
 influence on hard nuclear processes. The production of hidden and open charm at RHIC and LHC in the 
 more traditional framework of collinear factorization was discussed by R. Vogt  \cite{Vogt:2004hd}; in particular, she 
 examined the influence of several of the existing approaches to shadowing on the yields of charmed quarks.

\section{Hard probes of hot and dense QCD matter}

\subsection{Perturbative QCD -- the baseline}

No-one at present doubts the applicability of perturbative QCD to the description of "sufficiently" hard processes. 
Perturbative methods therefore provide a crucial baseline for the understanding of the attenuation of 
high momentum partons in hot and dense matter. 
Of particular interest to the participants was the long--standing 
puzzle of the apparent discrepancy between the yields of heavy quarks as measured at collider energies 
and the perturbative calculations. The problem, and possible solutions, was discussed by S. Frixione \cite{frixione}.  

\subsection{Jets and heavy quarks as a probe}

One of the most spectacular successes of the RHIC program is the discovery of the suppression of high 
transverse momentum particles, predicted as a signature of the quark--gluon plasma. An introduction 
to the problem, and an overview of the existing and future possibilities with the high momentum probes 
was given by X.-N. Wang \cite{wang}. 

\begin{figure} 
\resizebox{0.50\textwidth}{!}{%
\includegraphics[]{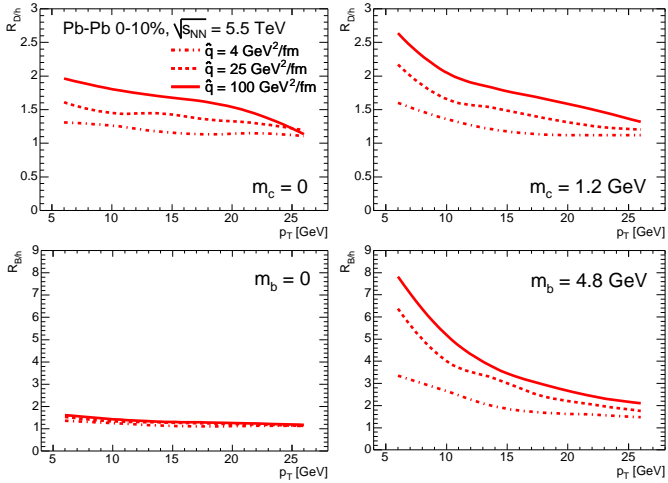}
}
\caption{The ratio of the nuclear modification factors in the production of heavy and light quarks in 
heavy ion collisions; from \cite{Wiedemann:2005gm,Armesto:2005iq}.}
\label{dead}
\end{figure}

The influence of the quark--gluon plasma on jet shapes and on the propagation of heavy quarks was the 
topic of U. Wiedemann's talk \cite{Wiedemann:2005gm}. The energy loss of heavy quarks was also the 
discussed by M. Djordjevic \cite{magdalena}. The results indicate a considerable enhancement in $D/\pi$ ratios (see Fig. \ref{dead})  
resulting from the interplay between the "dead cone effect" and the coherent multiple scattering, in qualitative agreement with other treatments \cite{Dokshitzer:2001zm}.

\begin{figure} 
\resizebox{0.50\textwidth}{!}{%
\includegraphics[]{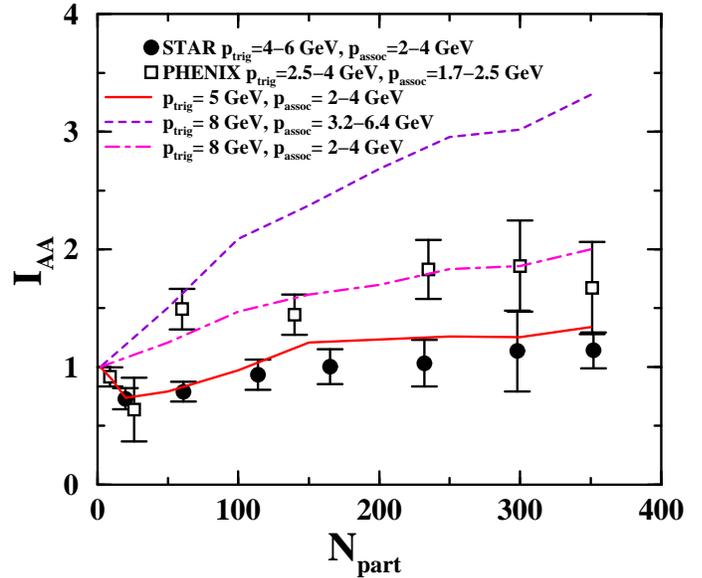}
}
\caption{The medium modification of associated hadron distribution from jet fragmentation 
in $Au-Au$ collisions at $\sqrt{s} = 200$ GeV for different choices of trigger $p_T$ and 
associated $p_T$; theory from is compared to STAR \cite{star-as} and PHENIX \cite{phenix-as} data. 
>From \cite{Majumder:2005ii,Majumder:2004pt}.}

\label{frag}
\end{figure}

A. Accardi \cite{Accardi:2005fu} investigated the relative importance of Cronin effect and jet quenching 
at different RHIC energies. An interesting analysis of di--hadron correlations in the fragmentation 
of the jets was presented by A. Majumder \cite{Majumder:2005ii}, who explored how the dense 
QCD matter affects the associated hadron distributions (see Fig. \ref{frag}).

\begin{figure} 
\resizebox{0.50\textwidth}{!}{%
\includegraphics[]{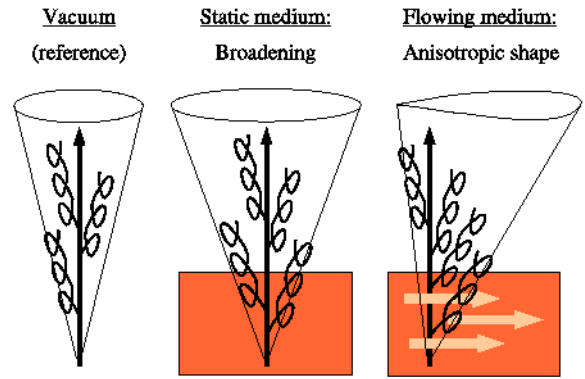}
}
\caption{The effect of collective flow on the jet shape; from \cite{Armesto:2005zn}.}
\label{flow}
\end{figure}

A novel effect of the influence of the hydrodynamical flow on the jet shape was considered by 
N. Armesto \cite{Armesto:2005zn}. He found that the flow can lead to an anisotropic jet shape, 
as illustrated in Fig. \ref{flow}.

The influence of the medium on the fragmentation of partons was also the topic of R. Hwa's talk \cite{Hwa:2005ay}. He suggested that because of the high density of partons in the quark--gluon plasma, 
the recombination of partons is a likely mechanism which can affect the composition and 
the transverse momentum distributions of the produced hadrons.  (For a related approach, see also \cite{Fries:2003kq}). 

\begin{figure} 
\rotatebox{270}{
\includegraphics[width=2.5in]{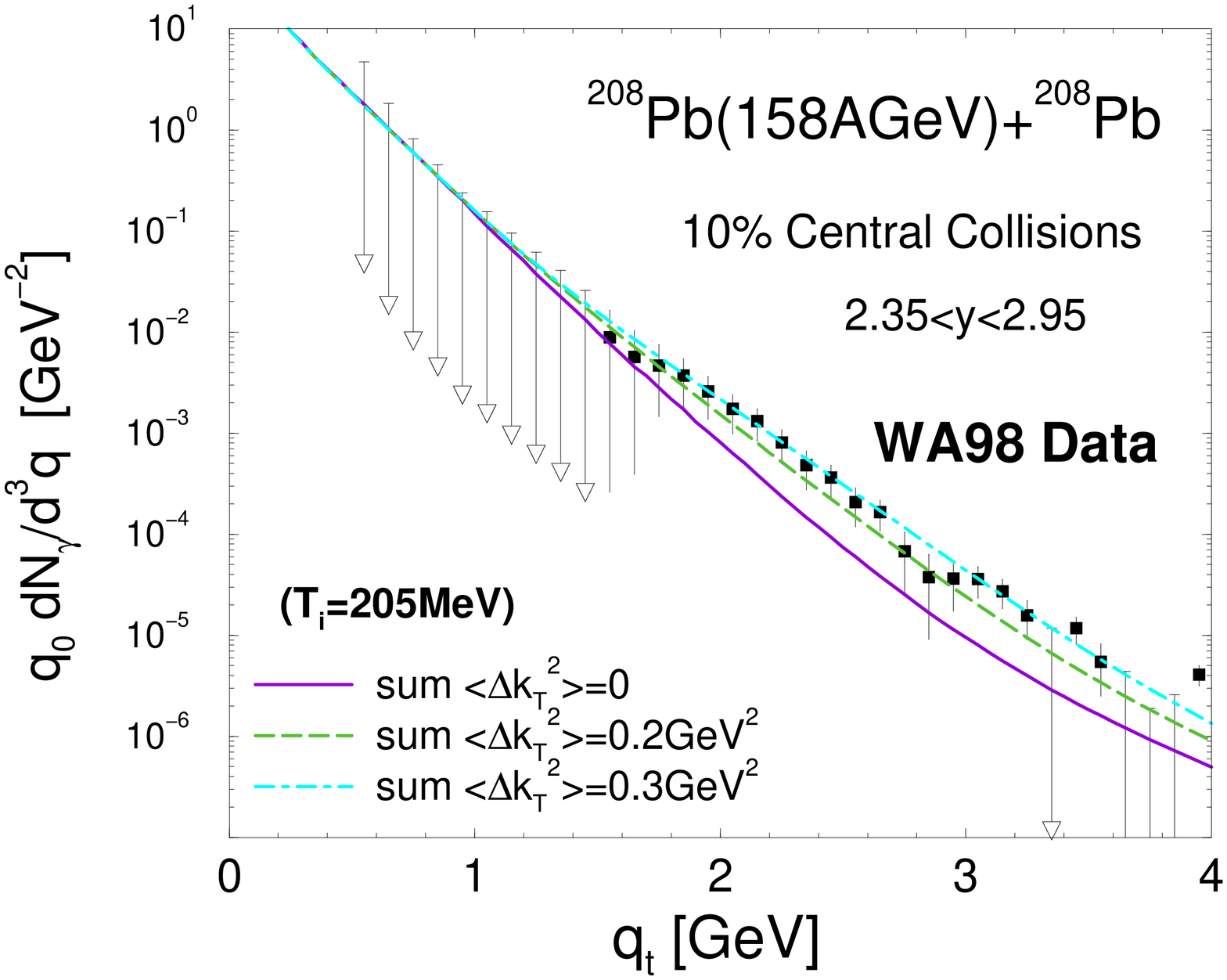}
}
\end{figure}
\begin{figure}
\rotatebox{270}{
\includegraphics[width=2.5in]{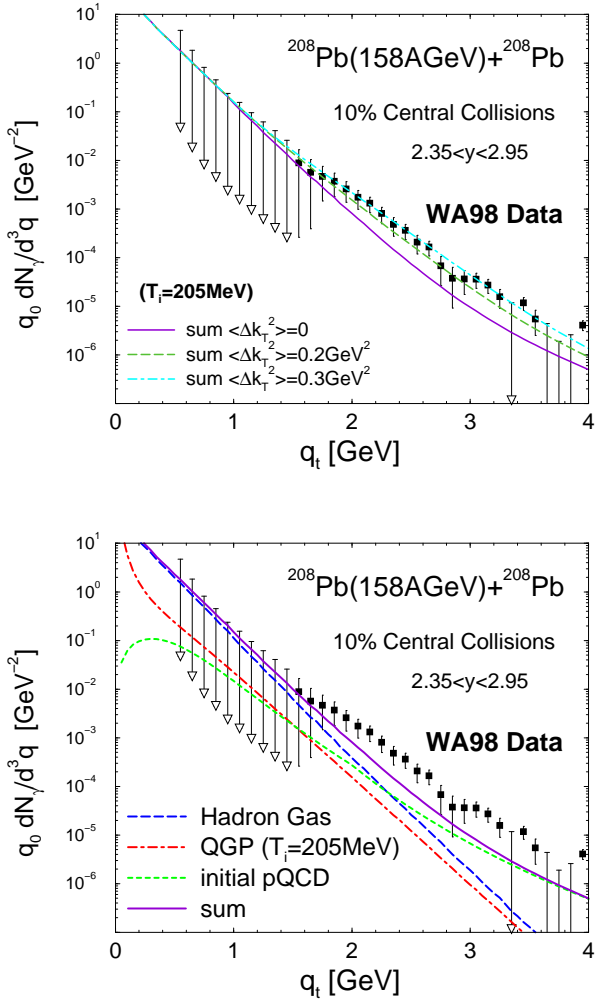}
}
\caption{The effect of the nuclear transverse momentum broadening on the measured photon spectrum (top panel) and the contributions of prompt and thermal spectra compared to the data on $Pb-Pb$ collisions 
at SPS from the WA98 Collaboration; from \cite{Gale:2005ri}.}
\label{dilept}
\end{figure}

\subsection{Electromagnetic probes}

The production of photons and dileptons from a hot quark-gluon matter remains a subject of vigorous 
theoretical and experimental studies. The state of the theoretical calculations has been reviewed 
at the Conference by C. Gale \cite{Gale:2005ri} and E. Shuryak. Gale emphasized that a variety 
of phenomena contribute to the photon and dilepton production, and they have to be carefully 
evaluated to make the extraction of the quark--gluon plasma component possible, see Fig. \ref{dilept}.

\section{Outlook}

This summary clearly does not capture the entirety  of the theoretical developments 
presented at the Conference -- it is impossible to fit the entire week of wonderful talks and exciting 
discussions in a few pages of written text. Nevertheless, I hope that a more complete picture 
can be reconstructed by looking at the original talks referenced here.
This is the picture of the field which is still at the very beginning -- prompted by the huge wave of new 
high quality data, the theorists are still in search of a coherent framework capable of describing 
the variety of the observed phenomena. 

However, in my opinion, the talks at the Conference show that 
we start to see the essential elements of this unified framework, and enough bright people with enough 
enthusiasm are working on the problem. Coupled with an impressive progress in experiment, 
this indicates that the ultimate goal of understanding QCD in the high temperature and strong field regimes 
may now be within reach.

\vskip0.2cm  
  
This work was supported by the U.S. Department of Energy under the contract DE-AC02-98CH10886.  
%
%

%
%

\end{document}